\documentclass[pra,twocolumn]{revtex4}%

\begin{document}
\title{No-Cloning and No-Deleting theorems through the existence of Incomparable states under LOCC}

\author{Amit Bhar, Indrani Chattopadhyay\protect\(
^{1}\protect \)
 and Debasis Sarkar\protect\(
^{2}\protect \)} \affiliation{Department of Applied Mathematics,
University of Calcutta, 92, A.P.C. Road, Kolkata- 700009, India}

\begin{abstract}
No-Cloning and No-Deleting theorems are verified with the constraint
on local state transformations via the existence of incomparable
states. Assuming the existence of exact cloning or deleting
operation defined on a minimum number of two arbitrary states, an
incomparable pair of states of the joint system between two parties
can be made to compare under deterministic LOCC. We have restricted
our proof with the assumption that the machine states of the cloning
or deleting operations do not keep any information about the input
states. We use the same setting to establish the no-cloning and
no-deleting theorems via incomparability that supports the
reciprocity of the two operations in their operational senses. The
work associates the impossibility of operations with the evolution
of an entangled system by LOCC.

PACS number(s): 03.67.Mn, 03.67.Hk


\end{abstract}

\maketitle

One of the most important task in quantum information processing is
to detect the allowable set of operations performed on quantum
systems. If someone wants to copy an arbitrary quantum information
encoded in a quantum state then no-cloning theorem \cite{wootters}
restricts one to copy arbitrary quantum information exactly. Quite
reverse to it, if we want to delete arbitrary quantum information
then we have a similar kind of restriction \cite{pati11, zurek}.
According to the no-deletion theorem \cite{pati11, zurek}, it is not
possible to delete arbitrary quantum information encoded in a
quantum state to a standard one. On the other hand, manipulation of
pure state entanglement provides us some other kind of restrictions
on the evolution of quantum systems. Sometimes a specific state may
be required to perform a specific information theoretic task. Then
Nielsen's criterion \cite{nielsen} determines the possibility of
inter-conversion of one pure entangled state shared between two
spatially separated parties to another by deterministic LOCC. This
result provides us a necessary and sufficient condition for
converting a bipartite pure entangled state to another by LOCC with
certainty. Now one may ask whether the no-go theorems and other
impossibilities only restrict the specific tasks or may be useful
for other kind of tasks that seems to be impossible otherwise? To
search for a common origin of these impossibilities one have to find
the possibility of interconnection between themselves within or
outside the quantum formalism. Here we provide a connection between
no-cloning and no-deleting theorems with the incomparability of pure
entangled states. The work shows, existence of either of the exact
cloning or deletion machine that act perfectly on any set of
non-orthogonal states, will imply local inter-conversion of a pair
of incomparable states with certainty.

We begin with some necessary background to our work. In quantum
information theory the no-go theorems are used to define intrinsic
properties of quantum systems beyond their usual status of imposing
restrictions over the systems. They allow quantum systems to perform
some computational tasks which are rather impossible by using
classical algorithms. In quantum cryptography \cite{cry}, the
possibility of detecting an eavesdropper having an access on the
communication channel emerges out of the well known no-cloning
\cite{wootters} theorem. In terms of information processing, cloning
can be viewed as the copying of information encoded in some systems
to other systems \cite{brus}. If $|\psi\rangle$ be the input state
then we describe exact cloning operation as $|\psi\rangle \otimes |b
\rangle \Rightarrow |\psi\rangle \otimes |\psi\rangle $, where $|b
\rangle$ is some suitably chosen blank state. Now quantum systems
will not provide complete accuracy of performing those operations on
arbitrary input states. Linearity of quantum operations establishes
precisely the impossibility of existence of an Universal Exact
Cloning Machine \cite{wootters,linear}. Unitarity of any quantum
evolution also shows that Universal Exact Cloning operation is not
physical in nature \cite{unitary}. Linearity of allowable quantum
operations further provides us another constraint which we termed as
No-Deletion theorem \cite{pati11,zurek}.  Deletion is quite a
reverse process than that of cloning. It is performed on two copies
of an arbitrary input state and is not possible to delete exactly
the information of one copy, keeping intact the information of the
other copy. In other words, the operation $|\psi\rangle \otimes
|\psi\rangle\Rightarrow |\psi\rangle \otimes |b\rangle$ is not
possible exactly for an arbitrary input state $|\psi\rangle$ with
certainty.

There are some other no-go theorems defined on single qubit systems,
such as the no-flipping theorem \cite{gisin}. From linearity of
quantum operations we find further the restriction of
no-partial-cloning, and other no-go theorems obtained from the
concepts of various quantum gates \cite{patili}. Efforts are made to
search the inter-relations between different no-go theorems and
relate them with other theories. For example, no-signaling principle
restricts any physical operation to evolve in such a way that can
not be used to send a signal faster than the speed of light.
No-signaling condition preserves all the impossibilities cited above
\cite{gisin1,signal2,signal3,flip}. Again, the constraint of
non-increase of entanglement under LOCC, described in a quite
similar way as that of the second law of thermodynamics. Applying
any local operations on the subsystems of a quantum system together
with classical communications between distant parties, it is
impossible to increase the entanglement of the joint system. The
no-cloning \cite{horo}, no-deleting \cite{rev}, no-flipping
\cite{flip} and many other impossibilities \cite{rev} are connected
with this constraint of information theory. Also the interrelation
between the cloning and flipping operations is revealed by the
conservation laws of simple classical theory \cite{Enk}. Now a very
new kind of information theoretic restriction on allowable quantum
operations observed through the existence of incomparable states
\cite{incompare}. This restriction retrieve the no-flipping theorem
\cite{incomfi} and also detects impossibility of some general
classes of local quantum operations \cite{impossible}. Here we want
to reveal a relation between this constraint with the very famous
no-cloning and no-deleting theorems. The work proceeds to verify the
reciprocity of the two no-principles by dealing them in a single
setting without verifying them separately. The connection between
all those no-go theorems of quantum systems with the impossibility
of inter-conversion of incomparable states would support the
existence of incomparable states beyond their mathematical status
from Nielsen's criterion. It provides also the nature of allowable
physical operations.

To present our work we need to define first the condition for a pair
of states to be incomparable with each other. The notion of
incomparability of a pair of bipartite pure entangled states is a
consequence of Nielsen's \cite{nielsen} majorization criterion.
Suppose we want to convert the pure bipartite state $|\Psi\rangle$
to $|\Phi\rangle$ shared between two parties, say, Alice and Bob by
deterministic LOCC. Consider the pair $(|\Psi\rangle$,
$|\Phi\rangle)$ in their Schmidt bases $\{|i_A\rangle ,|i_B\rangle
\}$ with decreasing order of Schmidt coefficients: $|\Psi\rangle=
\sum_{i=1}^{d} \sqrt{\alpha_{i}} |i_A i_B\rangle$, $|\Phi\rangle=
\sum_{i=1}^{d} \sqrt{\beta_{i}} |i_A i_B\rangle,$ where
$\alpha_{i}\geq \alpha_{i+1}\geq 0$ and $\beta_{i}\geq
\beta_{i+1}\geq0,$ for $i=1,2,\cdots,d-1,$ and $\sum_{i=1}^{d}
\alpha_{i} = 1 = \sum_{i=1}^{d} \beta_{i}$. The Schmidt vectors
corresponding to the states $|\Psi\rangle$ and $|\Phi\rangle$ are
$\lambda_\Psi\equiv(\alpha_1,\alpha_2,\cdots,\alpha_d)$ and
$\lambda_\Phi\equiv(\beta_1,\beta_2,\cdots,\beta_d)$. Then Nielsen's
criterion says $|\Psi\rangle\rightarrow| \Phi\rangle$ is possible
with certainty under LOCC if and only if $\lambda_\Psi$ is majorized
by $\lambda_\Phi,$ denoted by $\lambda_\Psi\prec\lambda_\Phi$ and
described as,
\begin{equation}
\begin{array}{lcl}\sum_{i=1}^{k}\alpha_{i}\leq
\sum_{i=1}^{k}\beta_{i}~ ~\forall~ ~k=1,2,\cdots,d
\end{array}
\end{equation}
It is interesting to note that as a consequence of non-increase of
entanglement by LOCC, if $|\Psi\rangle\rightarrow |\Phi\rangle$ is
possible under LOCC with certainty, then $E(|\Psi\rangle)\geq
E(|\Phi\rangle)$ [where $E(\cdot)$ denote the von-Neumann entropy of
the reduced density operator of any subsystem and known as the
entropy of entanglement]. Now in case of failure of the above
criterion (1), it is usually denoted by $|\Psi\rangle\not\rightarrow
|\Phi\rangle$. But it may happen that $|\Phi\rangle\rightarrow
|\Psi\rangle$ under LOCC. And if it happens that both
$|\Psi\rangle\not\rightarrow |\Phi\rangle$ and
$|\Phi\rangle\not\rightarrow |\Psi\rangle$ then we denote it by
$|\Psi\rangle\not\leftrightarrow |\Phi\rangle$ and describe
$(|\Psi\rangle, |\Phi\rangle),$ as a pair of incomparable states.
One of the peculiar feature of the existence of such incomparable
pairs is that we are really unable to say which state has a greater
amount of entanglement content than that of the other. For $2\times
2$ systems there are no pair of incomparable pure entangled states
as described above. For our purpose, we want to mention explicitly
the criterion of incomparability for a pair of pure entangled states
$|\Psi\rangle, |\Phi\rangle$ of $m\times n$ system where $\min \{
m,n \}=3$. Suppose the Schmidt vectors corresponding to the two
states are $(a_1, a_2, a_3)$ and $(b_1, b_2, b_3)$ respectively,
where $a_1> a_2> a_3>0~,~b_1> b_2> b_3>0~,~a_1+ a_2+ a_3=1=b_1+ b_2+
b_3$. Then it follows from Nielsen's criterion that $|\Psi\rangle,
|\Phi\rangle$ are incomparable \cite{incompare} if and only if,
either of the pair of relations
\begin{equation}
\begin{array}{lcl}
a_1 > b_1 ~~\verb"&"~~ a_3 > b_3\\
 b_1 > a_1 ~~\verb"&"~~ b_3 > a_3
\end{array}
\end{equation} will hold.

Our paper concerns with one-to-two copy exact cloning operation on a
minimum number of two arbitrary states $|0\rangle,|\psi \rangle$ in
the following form
\begin{equation}
\begin{array}{lcl}
& &|0\rangle|b\rangle \longrightarrow |0\rangle|0\rangle\\
& &|\psi \rangle |b\rangle \longrightarrow |\psi\rangle|\psi\rangle
\end{array}
\end{equation}
where $|b\rangle$ is a suitably chosen blank state.

We concentrate entirely within the quantum formalism and for that
reason we assume the machine states do not keep any information
about the input states. So we drop the machine states in the
definition of the cloning operation. No-cloning theorem then turns
out to be the impossibility of this operation for arbitrary state
$|\psi \rangle$.

Now we consider that Alice and Bob, two spatially separated parties
have a particular setting of a pure bipartite state in the form
given below
\begin{equation}
\begin{array}{lcl}
|\Omega^i\rangle_{AB}&=& \frac{1}{\sqrt{N^i}} \{ |1\rangle_A |0 \psi
0 \psi + \psi 0 \psi 0\rangle_B  + |2 \rangle_A |0 \psi \psi 0 \\ &
& - \psi 0 0 \psi\rangle_B  + |3 \rangle _A |0 0 \psi \psi - \psi
\psi 0 0 \rangle_B \} \otimes |b\rangle_B
\end{array}
\end{equation}

This is a six particle state where Alice has one qutrit and Bob has
four qubits entangled with Alice's system together with a separate
qubit in the form of blank state $|b\rangle$. So the joint system is
of $3\times32$ dimension, where $N^i=2(3-{\alpha}^4)$ be the
normalizing constant and $|\psi \rangle= \alpha |0\rangle + \beta
|1\rangle$ be an arbitrary qubit with $|\alpha|^2 +|\beta|^2=1$. As
the arbitrary input state $|\psi \rangle$ can be written in the form
$|\psi \rangle= \cos \frac{\theta}{2}|0\rangle + e^{-i \phi}\sin
\frac{\theta}{2} |1\rangle$, where $\theta, \phi$ satisfy the
following equations $0 \leq \phi \leq 2\pi$, $-\frac{\pi}{2} \leq
\theta \leq \frac{\pi}{2}$, hence without loss of generality, the
parameter $\alpha$ is treated here as a real constant.

Tracing out Bob's local system we compute the initial reduced
density matrix $\rho^i _A$ on Alice's side in the following form
\begin{equation}
\begin{array}{lcl}
\rho^i _A &=& tr_B [~|\Omega^i \rangle_{AB} \langle \Omega^i|~]\\
&=& \frac{1}{N^i}~ \{2(1+ | \alpha |^4 )P[|1\rangle]+ 2(1- | \alpha
|^4 )\\ & &(P[|2\rangle]+P[|3\rangle]) \}
\end{array}
\end{equation}
where $P[|j\rangle]= |j\rangle \langle j|,$ for any $j$. The Schmidt
vector of the initial state can be written as $ \lambda^i =
({\lambda^i}_1, {\lambda^i}_2, {\lambda^i}_2)$ where $\lambda^i_1 =
\frac{1+{\alpha}^4}{3-{\alpha}^4}$ and $\lambda^i_2 =
\frac{1-{\alpha}^4}{3-{\alpha}^4}$. Hence $\lambda^i_{max} = \max
\{\lambda^i _1, \lambda^i_2\}= \lambda^i _1$ and $\lambda^i_{min} =
\min \{\lambda^i _1, \lambda^i_2\}= \lambda^i _2$. If the cloning
operation defined in equation$(3)$ exists and is applied on Bob's
local system (say on his fourth qubit together with the blank
state), the joint pure state shared between Alice and Bob could be
exactly transformed to the pure state,
\begin{equation}
\begin{array}{lcl}
|\Omega^f \rangle_{AB} &=& \frac{1}{\sqrt{N^f}}~ \{|1\rangle_A |0
\psi 0 \psi \psi + \psi 0 \psi 0 0\rangle_B + |2 \rangle_A  |0 \psi
\psi 0 0  \\ & & - \psi 0 0 \psi \psi \rangle_B  + |3 \rangle _A |0
0 \psi \psi \psi  - \psi \psi 0 0 0\rangle_B \}
\end{array}
\end{equation}
where $N^f=2(3-{\alpha}^5)$ be the normalizing constant. The final
reduced density matrix on Alice's side would be
\begin{equation}
\begin{array}{lcl}
\rho^f _A &=& tr_B [~ |\Omega^f \rangle_{AB} \langle \Omega^f|~ ] \\
&=& \frac{1}{N^f}~ \{2(1+  \alpha ^5 )P[|1\rangle]+ 2(1- \alpha ^5
)(P[|2\rangle]+P[|3\rangle]) \\& &~ ~ ~ ~ - 2 \alpha ^2 (1-
\alpha)(|2\rangle \langle3| + |3\rangle \langle2| ) \}
\end{array}
\end{equation}
Hence the Schmidt coefficients of the final state $\rho^f_A $ are $
\{ \frac{1+{\alpha}^5}{3-{\alpha}^5} , \frac{(1 + {\alpha}^2)(1-
{\alpha}^3)}{3-{\alpha}^5} , \frac{(1 - {\alpha}^2)(1+
{\alpha}^3)}{3-{\alpha}^5} \}$. If we denote $\lambda^f_1=
\frac{1+{\alpha}^5}{3-{\alpha}^5}$ , $\lambda^f_2 = \frac{(1 +
{\alpha}^2)(1- {\alpha}^3)}{3-{\alpha}^5}$ and $\lambda^f_3 =
\frac{(1 - {\alpha}^2)(1+ {\alpha}^3)}{3-{\alpha}^5},$  then
$\lambda^f_{min} = \min \{\lambda^f _1, \lambda^f_2, \lambda^f_3 \}
= \lambda^f_3 $  and thus $\lambda^f_{max} = \max \{\lambda^f _1,
\lambda^f_2\}$. Now using simple algebra we find,
$\lambda^f_1<\lambda^i_1$ and also $\lambda^f_2<\lambda^i_1$ (if,
$\lambda^f_1<\lambda^f_2$), which implies that,
$\lambda^f_{max}<\lambda^i_{max}$. Finally we get, $\lambda^f_{min}
=\lambda^f_3<\lambda^i_2 = \lambda^i_{min}.$ These inequalities
clearly indicate the nature of incomparability of the pair of pure
bipartite states $| \Omega^i \rangle$ and $| \Omega^f \rangle $. The
incomparability of the states imply that the final state $| \Omega^f
\rangle $ can not be achieved from the initial state $| \Omega^i
\rangle$ through LOCC with certainty. Thus we are compelled to
conclude that the cloning operation performed on Bob's local system
to implement the transformation $| \Omega^i \rangle \rightarrow |
\Omega^f \rangle $ locally, is not a physical operation. In other
words the exact cloning operation is not possible, for any pair of
arbitrary non-orthogonal input states. This ensures the successful
establishment of no-cloning theorem.

Now if we further treat $| \Omega^f \rangle $ as the initial pure
bipartite state, shared between Alice and Bob and assume the
existence of an exact deleting machine again defined on only two
arbitrary input qubit $ |0\rangle,| \psi \rangle$ as
\begin{equation}
\begin{array}{lcl}
& &|0\rangle|0\rangle ~\longrightarrow |0\rangle|b \rangle \\
& &|\psi\rangle |\psi \rangle \longrightarrow |\psi \rangle
|b\rangle
\end{array}
\end{equation}
and apply this machine on Bob's local system the joint state
$|\Omega^f \rangle$ between them can be converted into the state
$|\Omega^i \rangle$. Under the previous arguments it could be easily
proved that $| \Omega^i \rangle \not \leftrightarrow | \Omega^f
\rangle $, i.e., the transformation $| \Omega^f \rangle \rightarrow
| \Omega^i \rangle $ is impossible by LOCC with certainty. This
impossibility directly indicates that the deleting operation defined
in equation$(8)$ is not a valid physical operation for arbitrary
input states. So this leads us to the formal no-deleting theorem.

In conclusion this work connects the two famous no-go theorems from
a new viewpoint that restricts the possible evolution of any quantum
system through local operations. It shows the physical reason behind
the existence of ``\emph{Incomparable Pair of Pure Bipartite
States}". This connection makes a bridge between two different
aspects of information processing theory. Moreover the most
interesting feature is that the no-cloning and no-deleting theorems
are treated in the same platform and thus we see the reciprocity of
the two theorems from an operational point of view. Although for
simplicity we assume that the machine states for both the cloning
and deleting operations do not contain any information about the
input qubit state, one may not assume this restriction. The result
also holds if we consider the general scenario. Another interesting
part in our proof is that the state we have considered, has a
peculiar kind of symmetry and we require $3\times 32$ dimensional
system to prove our result. However one may search for a proof in
lower dimensional systems. \\

{\bf Acknowledgement.} The authors thank the referee for his/her
valuable comments and suggestions. I.C. also acknowledges CSIR,
India for providing fellowship during this work.


 {\protect\( ^{1}\protect \)ichattopadhyay@yahoo.co.in} {\protect\( ^{2}\protect
\)dsappmath@caluniv.ac.in}

\end{document}